\documentclass[twocolumn,showpacs,preprintnumbers,draftclsnofoot]{revtex4}
\usepackage{amsmath}
\usepackage{amssymb}
\usepackage{graphicx}
\usepackage{dcolumn}
\usepackage{bm}
\usepackage{amssymb}
\usepackage{graphicx}
\usepackage{dcolumn}
\usepackage{bm}

\begin{document}

\title{ Tuning of Bilayer Graphene Heterostructure by Horizontally Incident Circular Polarized Light  }
\author{Ma Luo\footnote{Corresponding author:luoma@gpnu.edu.cn} }
\affiliation{School of Optoelectronic Engineering, Guangdong Polytechnic Normal University, Guangzhou 510665, China}

\begin{abstract}

We theoretically investigated the Floquet states of bilayer graphene heterostructure under the irradiation by horizontally incident circular polarized light. The in-plane and out-of-plane electric field of the light periodically perturbs the intra-layer and inter-layer hopping, respectively. For circular polarized light, the two components of the electric field has $\pi/2$ phase difference, so that the two types of hopping are periodically perturbed with the $\pi/2$ phase difference, which modify the effective inter-layer hopping. We focus on the model of bilayer graphene in the heterostructure of antiferromagnetic van der Walls spin valve. The amplitude of the irradiation can tune the band gap and topological properties of the bulk state. The spin-polarized quantum anomalous Hall phase with Chern number being one is predicted. The incident angle of the irradiation can tune the band gap and dispersion of the edge states in zigzag nanoribbons.

\end{abstract}

\pacs{00.00.00, 00.00.00, 00.00.00, 00.00.00}
\maketitle

\section{Introduction}

Van der Walls heterostructures consisting of graphene and varying types of substrates are outstanding candidate as spintronic devices, which have been  extensively investigated \cite{Francois20,YuanLiu16,Xiuqiang20,Xuechao20,luo19}. Proximity effect between graphene and the substrates induces spin-dependent interactions, including intrinsic and Rashba spin-orbit coupling (SOC) \cite{Rashba09,Zhenhua10,Jayakumar14}, ferromagnetic and antiferromagnetic exchange field in the graphene layer \cite{Zollner20,PetraHogl20,Sushant19,Frank18,Offidani17,Gmitra16,Morpurgo15,Gmitra15}. One of the most attractive heterostructure is Van der Walls spin valves, which are consisted of bilayer graphene (BLG) being sandwiched between two substrates \cite{Cardoso18,luo19}. Flipping the exchange field of one substrate opens or closes the bulk band gap. In addition, the vertical gated voltage can control the band gap as well. In the presence of SOC, the BLGs are driven into varying topological states, such as quantum spin Hall (QSH), quantum valley Hall (QVH) and quantum anomalous Hall (QAH) phases \cite{ZhenhuaQiao10,MotohikoEzawa12,Offidani18,Yafei17,Abdulrhman18,maluo18,maluo19v,maluo19f}. By tuning the topological phase, the edge states are controlled.

On the other hand, periodic perturbation of graphene drives the quantum system into a dynamic state, which is described by the Floquet theory \cite{Rodriguez08,Takashi09,Savelev11,ArijitKundu14,Taboada17,Taboada171,DalLago15,MichaelVogl19,MichaelVogl20}. The electronic structure of the Floquet states can be measured by applying the experimental method of angle-resolved photoemission spectroscopy (ARPES) \cite{YHWang13,Farrell16,YaoWang18,HaifengYang18,BaiqingLv19}. If the perturbation is due to irradiation by normally incident circular optical field, the time reversal symmetric is broken, so that the system could be driven into the QAH phase or Floquet topological insulator phase \cite{Inoue10,Kibis10,Calvo11,ShuTing16,Mukherjee18,Yunhua17,Ledwith18,HangLiu18,LongwenZhou18,maluo21F}. The Floquet topological phase is featured by the presence of chiral edge states in graphene nanoribbon \cite{Piskunow14,Usaj14,Claassen16,Tahir16,Puviani17,Hockendorf18,PerezPiskunow15}, which enable robust topological transport through the nanoribbon \cite{AaronFarrell15,AaronFarrell16}. The topological edge transport can be measured by on-chip photoconductive device that is driven by femtosecond optical pulse \cite{JWMcIver20}. Combination of optical driven and external magnetic field induces valley polarized state in BLG \cite{Abergel11}. Recently, varying types of topological flat band in optically driven twisted BLG have been proposed, which provide more candidates of strongly correlated topological phases \cite{YantaoLi20,MichaelVogl20t,MartinRodriguez21,DanteKennes21}. In the high frequency approximation, the effect of the irradiation can be described by Haldane mass term \cite{Haldane88,Kitagawa11}. The presence of the Haldane mass term in the ferromagnetic van der Walls spin valve induces spin-valley polarized QAH phase in the BLG \cite{Xuechao20}. Similar effect could be engineered in the antiferromagnetic van der Walls spin valve.

In this paper, we seek additional mechanism to tune the antiferromagnetic van der Walls spin valve, which is irradiated  by horizontally incident circular polarized optical field. The tuning parameters of the opto-spintronic system include the incident angle as well as the polarization and intensity of the optical field. For monolayer graphene, irradiation by horizontally incident circular optical field has trivial effect that is similar to the irradiation by linear polarized optical field, because the out-of-plane electric field has negligible effect on the intra-layer hopping. By contrast, the inter-layer hopping between the two graphene layers in the BLG is perturbed by the out-of-plane electric field \cite{RodriguezReview,Rodriguez20,Rodriguez20a,Rodriguez21}, so that the irradiation can tune the band structure with non-trivial effect. We theoretically studied the tuning of the antiferromagnetic van der Walls spin valves, which is BLG with antiferromagnetic exchange field, staggered sublattice potential, vertical gated voltage and (or) intrinsic SOC. The Rashba SOC due to proximity effect is relatively small, which is neglected in our model. We focused on the BLG with Bernal stacking order, which is commonly found in experimental sample. The bulk band structures and topological phases are modeled by the high frequency approximation on the Dirac Fermion model \cite{Kitagawa11,Goldman14,Grushin14}. We found that the irradiation with sufficient intensity induces a topological phase transition for each band valley. The band structures of the zigzag nanoribbons are modeled by tight binding model \cite{Shirley65,Sambe73,Kohler05}. By engineering the materials and optical parameters, the nanoribbon with one pair of spin polarized chiral edge states with van-Hove singularity is designed.

The article is organized as follows: In Sec. II, the high frequency expansion of the Dirac Fermion model is applied to studied the tuning of bulk band gap and topological phase by the irradiation. In Sec. III, the tight binding model is applied to studied the topological edge state of the zigzag nanoribbon under the irradiation. In Sec. IV, the conclusion is given.

\section{Tuning of the bulk band structure}

The incident circular polarized optical field propagates along the in-plane direction. The angle between the wave vector of the optical field and the $\hat{x}$ axis is $\theta$. The optical frequency is assumed to be $\hbar\Omega=7$ eV, with the corresponding wavelength being 177 nm. Because the wave length of the optical field is much larger than the lattice constant of graphene, the spatial dependent of the optical field is neglected. Thus, the optical field is described by oscillation of the vector potential, $\mathbf{A}(t)=\hat{x}A_{x}\sin(\Omega t)+\hat{y}A_{y}\sin(\Omega t)+\hat{z}A_{z}\cos(\Omega t)$. The in-plane component of the vector potential is designated as $\mathbf{A}_{\|}(t)=\hat{x}A_{x}\sin(\Omega t)+\hat{y}A_{y}\sin(\Omega t)$. The amplitude of the in-plane vector potential is designated as $A_{r}$, where $A_{x}=-A_{r}\sin(\theta)$ and $A_{y}=A_{r}\cos(\theta)$. The left and right circular polarized optical field have $A_{r}=\eta A_{z}$ with $\eta=\pm1$.  The in-plane and out-of-plane component of the vector potential have the same amplitude, but have $\eta\pi/2$ phase difference. The amplitude of the electric field is $E_{0}=|A_{r}|/\Omega=|A_{z}|/\Omega$.



\subsection{Effective Hamiltonian of Dirac Fermion model}

Because the optical frequency is larger than the bandwidth of graphene, the band structure of BLG near to the Fermi level could be approximated by the Dirac Fermion model. For the BLG with Bernal (AB) stacking order and effective mass terms $\Delta_{+}$ and $\Delta_{-}$ at top and bottom layer, the Hamiltonian of the Dirac Fermion model is
\begin{equation}
H=\left[\begin{array}{cccc}
V+\Delta_{+} & \hbar v_{F}k^{\tau}_{-} & 0 & 0 \\
\hbar v_{F}k^{\tau}_{+} & V-\Delta_{+} & t_{\bot}^{+} & 0 \\
0 & t_{\bot}^{-} & -V+\Delta_{-} & \hbar v_{F}k^{\tau}_{-} \\
0 & 0 & \hbar v_{F}k^{\tau}_{+} & -V-\Delta_{-} \\
\end{array}\right]\label{HamiltonianD1}
\end{equation}
where $k^{\tau}_{\pm}=\tau k_{x}\pm ik_{y}$ with $\tau=\pm1$ being the valley index of the K and K$^{\prime}$ valleys, $k_{x}$ and $k_{y}$ are the wave number relative to the K or K$^{\prime}$ point, $2V$ is the potential difference between the top and bottom layers due to the gated voltage, $t_{\bot}^{+}=t_{\bot}^{-}=-0.39$ eV is the interlayer hopping constant, and $v_{F}$ is the Fermi velocity of graphene. The effective mass term is consisted of three components, as $\Delta_{\varsigma}=\lambda_{\Delta}^{\varsigma}+\lambda_{AF}^{\varsigma}\hat{s}_{z}+\lambda_{I}^{\varsigma}\tau\hat{s}_{z}$, where $\varsigma=\pm1$ represents the top and bottom layer.  The first term is the staggered sublattice potential. The second term is due to the presence of antiferromagnetic exchange field, where $\hat{s}_{z}=\pm1$ is the spin index. The third term is due to the presence of intrinsic SOC. The physical properties of each spin and band valley can be revealed by studying the model in Eq. (\ref{HamiltonianD1}).


In the presence of the irradiation, the Hamiltonian become time-dependent, which can be modeled by time-independent effective Hamiltonian. Because the Bloch states are spatially periodic along the in-plane direction, the dipolar approximation can be applied for the intra-layer hopping terms in the time-dependent Hamiltonian. Applying the the Peierls substitution for the in-plane optical field, $\mathbf{k}$ is replaced by $\mathbf{k}+ e_{0}\mathbf{A}_{\|}(t)/\hbar$ in the time-dependent Hamiltonian. On the other hand, the Bloch states are spatially localized along the out-of-plane direction, so that the inter-layer hopping terms in the time-dependent Hamiltonian are described by non-perturbative formula. The out-of-plane optical field induces a time-dependent Peierls phases \cite{Peierls33}, so that $t_{\bot}^{\pm}$ is replaced by $t_{\bot}^{\pm}e^{\frac{\pm i2\pi d_{z}A_{z}\cos(\Omega t)}{\Phi_{0}}}=t_{\bot}^{\pm}\sum_{m=-\infty}^{+\infty}{i^{m}J_{m}(\pm2\pi d_{z}A_{z}/\Phi_{0})e^{im\Omega t}}$, where $\Phi_{0}=\pi\hbar/e_{0}$ is the magnetic flux quantum, $d_{z}$ is the distance between the two graphene layers \cite{Calvo13}. As a result, the time-dependent Hamiltonian can be expanded as $H(t)=\sum_{m=-\infty}^{+\infty}{H_{m}e^{im\Omega t}}$. The effective Hamiltonian is given as $H^{eff}=H_{0}+\sum_{m>0}{\frac{[H_{+m},H_{-m}]}{m\Omega}}+O(\frac{1}{\Omega^{2}})$. Applying the high frequency expansion to the Hamiltonian (\ref{HamiltonianD1}), the effective Hamiltonian is given as
\begin{equation}
H^{eff}=\left[\begin{array}{cccc}
V+\Delta_{+} & \hbar v_{F}k^{\tau}_{-} & t_{\bot,1} & 0 \\
\hbar v_{F}k^{\tau}_{+} & V-\Delta_{+} & t_{\bot,0} & -t_{\bot,1} \\
t_{\bot,1}^{*} & t_{\bot,0} & -V+\Delta_{-} & \hbar v_{F}k^{\tau}_{-} \\
0 & -t_{\bot,1}^{*} & \hbar v_{F}k^{\tau}_{+} & -V-\Delta_{-} \\
\end{array}\right]\label{HamiltonianD1e}
\end{equation}
where $t_{\bot,0}=t_{\bot}J_{0}(2\pi d_{z}A_{z}/\Phi_{0})$, $t_{\bot,1}=2\hbar v_{F}(\tau A_{x}+iA_{y})t_{\bot}J_{1}(2\pi d_{z}A_{z}/\Phi_{0})/(\hbar\Omega)$. Thus, the irradiation effectively change the inter-layer hopping in two way: firstly, the vertical hopping strength between the nearest neighbor inter-layer sites is renormalized; secondly, the quantum interfere between the in-plane and out-of-plane perturbation generates the next nearest neighbor inter-layer hopping. The band structure and wave function of the Floquet states can be obtained by diagonalization of Eq. (\ref{HamiltonianD1e}). If the BLG has AA stacking order, the Hamiltonian of the Dirac Fermion model is
\begin{equation}
H=\left[\begin{array}{cccc}
V+\Delta_{+} & \hbar v_{F}k^{\tau}_{-} & t_{\bot}^{+} & 0 \\
\hbar v_{F}k^{\tau}_{+} & V-\Delta_{+} & 0 & t_{\bot}^{+} \\
t_{\bot}^{-} & 0 & -V+\Delta_{-} & \hbar v_{F}k^{\tau}_{-} \\
0 & t_{\bot}^{-} & \hbar v_{F}k^{\tau}_{+} & -V-\Delta_{-} \\
\end{array}\right]\label{HamiltonianD1AA}
\end{equation}
Applying the high frequency expansion, one can find that the effective Hamiltonian is the same as the unperturbed Hamiltonian, because the commutations $[H_{+m},H_{-m}]$ are equal to zero. As a result, the quantum state of the BLG with AA stacking order is not modified by the horizontally incident circular polarized optical field.


The topological feature of the Dirac Fermion band is characterized by the Chern number, which can be obtained by integrating the Berry curvature over the momentum space. The Berry curvature at each momentum is calculated as
\begin{equation}
\mathfrak{B}_{n}(\mathbf{k})=-\sum_{n\ne n^{\prime}}{\frac{2Im\langle\psi_{n\mathbf{k}}|v_{x}|\psi_{n^{\prime}\mathbf{k}}\rangle\langle\psi_{n^{\prime}\mathbf{k}}|v_{y}|\psi_{n\mathbf{k}}\rangle}{(\varepsilon_{n^{\prime}\mathbf{k}}-\varepsilon_{n\mathbf{k}})^{2}}}
\end{equation}
, where $n$ is band index of the eigenstates $|\psi_{n\mathbf{k}}\rangle$ with eigenvalue $\varepsilon_{n\mathbf{k}}$, $v_{x(y)}$ are velocity operator \cite{PerezPiskunow15,Rudner13,Kitagawa10}. The Chern number of each valley and spin is designated as $\mathcal{C}_{s_{z}}^{\tau}$. In the absence of the irradiation or intrinsic SOC, the BLG is in the QVH phase, because the Chern numbers of opposite band valleys are opposite in sign. The valley Chern number is defined as the difference between the Chern number of the two band valleys, i.e. $\mathcal{C}_{s_{z}}^{+}-\mathcal{C}_{s_{z}}^{-}$. The valley Chern number could be positive or negative two. The phase boundary between the two QVH phase with opposite valley Chern number is given as $2V=-\Delta_{+}-\Delta_{-}$. In the presence of intrinsic SOC, the BLG could be driven into quantum spin Hall (QSH) phase or spin-polarized quantum anomalous Hall (QAH) phases, by engineering the combination of ($V$, $\lambda_{\Delta}^{\varsigma}$, $\lambda_{AF}^{\varsigma}$, $\lambda_{I}^{\varsigma}$). In the presence of irradiation, the Chern number of each valley and spin could be tuned to zero, as shown in the following subsection.

\subsection{Phase diagram }

Counting the spin and valley indices, there are four band valleys for the BLG. We firstly studied the idealized antiferromagnetic spin valves, where $\lambda_{\Delta}^{\varsigma}$ and $\lambda_{I}^{\varsigma}$ are zero, so that $\Delta_{\pm}=\lambda_{AF}^{\pm}$.

\begin{figure}[tbp]
\scalebox{0.58}{\includegraphics{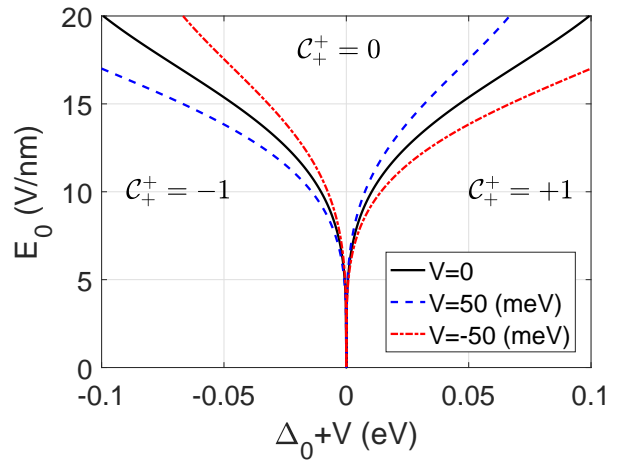}}
\caption{ The phase diagram of the irradiated spin valve with $\Delta_{+}=\Delta_{-}\equiv\Delta_{0}$, $\lambda_{\Delta}^{\varsigma}=0$ and $\lambda_{I}^{\varsigma}=0$. The phase boundary with $V=0$, $V=50$ meV, and $V=-50$ meV are plotted as solid (black), blue (dashed) and red (dash-dotted) lines, respectively. The Chern number of the $\tau=+1$ valley is marked in each phase regime. }
\label{fig_DiracFloquetBand}
\end{figure}

For the system with $\Delta_{+}=\Delta_{-}\equiv\Delta_{0}$ and $V\ne-\Delta_{0}$, the BLG have finite gap, which corresponds to the OFF state of the spin valve.  For these systems, the band structures of two valleys and spins are the same, so that we focus on the band structure of spin $s_{z}=+1$ and valley $\tau=+1$ to study the gap closing condition. Because the particle-hole symmetric is preserved, the Fermi level is at zero. In the presence of irradiation, as the amplitude of the optical field increases, the band gap is decreased, closed and reopened. The analytical solution of the condition of gap closing is obtained by diagonalization of the Hamiltonian in Eq. (\ref{HamiltonianD1e}). Since the Hamiltonian preserve the particle-hole symmetric, the gap closes at energy zero. However, the optical irradiation breaks the in-plane rotational symmetric, so that the gap closing does not occur at the K or K$^{\prime}$ point, but beyond the K or K$^{\prime}$ point with $k_{x(y)}\ne0$. The solution that two energy levels equate to zero occurs at the momentum point $(k_{x},k_{y})$ with $k_{x}=k_{r}\cos(k_{\phi})$, $k_{y}=k_{r}\sin(k_{\phi})$, $k_{r}=-2iA_{r}t_{\bot}J_{1}(2\pi d_{z}A_{z}/\Phi_{0})/(\hbar\Omega)\pm\sqrt{-\Delta_{0}^{2}\pm it_{\bot}J_{0}(2\pi d_{z}A_{z}/\Phi_{0})(\Delta_{0}+V)+V^2}/(\hbar v_{F})$ and $k_{\phi}=\theta-\pi$. At the critical amplitude of the optical field $(A_{r},A_{z})$ that $k_{r}$ is real, the gap is closed. Thus, the critical condition is obtained by numerically finding the zero point of $Im[k_{r}]=0$ for a given $(A_{r},A_{z})$. The critical amplitude of the optical field is dependent on both $\Delta_{0}$ and $V$. The numerical results of the critical value versus $\Delta_{0}+V$ for different $V$ are plotted in Fig. \ref{fig_DiracFloquetBand}. After the amplitude of the optical field exceed the critical value and the gap is reopened, the Chern number of the Dirac Fermion band  become zero.  As a result, the system is driven from QVH phase into a topologically trivial phase. The Chern number of spin $s_{z}=+1$ and valley $\tau=+1$ in each phase regime is marked in Fig. \ref{fig_DiracFloquetBand}.

\begin{figure}[tbp]
\scalebox{0.58}{\includegraphics{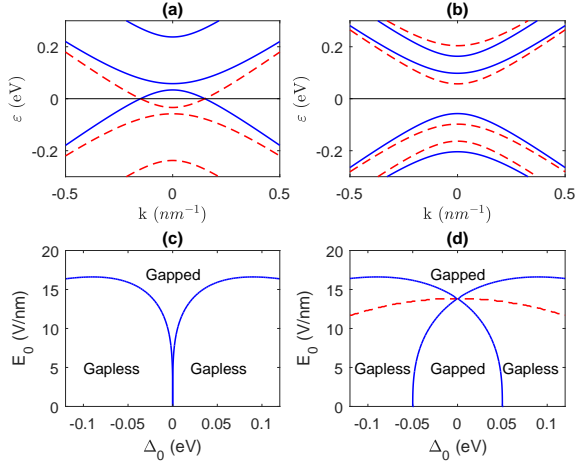}}
\caption{ (a) and (b) are band structure of the BLG with $\Delta_{+}=-\Delta_{-}=50$ meV, $E_{0}=12$ $V/nm$ and $E_{0}=20$ $V/nm$, respectively. The other parameters are $V=0$, $\lambda_{\Delta}^{\varsigma}=0$ and $\lambda_{I}^{\varsigma}=0$. The band structure of spin up and down are plotted as blue (solid) and red (dashed) lines, respectively. (c) The boundary between the phase regime with zero and finite band gap. (d) As the gate voltage being changed to $V=50$ meV, the boundaries between the phase regime with zero and finite band gap are plotted as blue (solid) lines. The phase boundary at which the direct band gap of each spin and valley closes is plotted as red (dashed) line. }
\label{fig_DiracFloquetBandAnti}
\end{figure}

For the system with $\Delta_{+}=-\Delta_{-}=\Delta_{0}$ and $|\Delta_{0}|>V$, the BLG is gapless, which corresponds to the ON state of the spin valve. For these systems, the band structures of the two valleys are the same; but those of the two spins are different. For the system with $V=0$ and $E_{0}=0$, the direct band gap of each spin is zero. Turning on the irradiation open a direct band gap for each spin. Because the particle-hole symmetric in this system is broken, the direct band gaps of two spins have different energy range. Thus, the band structure of the whole system is gapless, as shown in Fig. \ref{fig_DiracFloquetBandAnti}(a). As the amplitude of the irradiation increases, the direct band gaps of two spins become larger, and have overlapping energy range, so that the band gap of the whole system is opened, as shown in Fig. \ref{fig_DiracFloquetBandAnti}(b). The boundary between the phase regime with zero and finite gap is numerically calculated and plotted in Fig. \ref{fig_DiracFloquetBandAnti}(c). For the system with $V\ne0$ and $E_{0}=0$, the direct band gap of each spin is nonzero. Turning on the irradiation firstly decreases and closes the direct band gap, and then reopen the direct band gap. The critical value of the gap closing is plotted as red (dashed) line in Fig. \ref{fig_DiracFloquetBandAnti}(d). After the gap being reopened, the Chern number of each valley and spin is switch to zero. Meanwhile, the band gap of the whole system is tuned. As the energy ranges of the band gap of two spins overlap, the band gap of the whole system is opened. The corresponding phase boundaries of the transition of the direct band gap (the whole band gap) are plotted as blue (solid) lines in Fig. \ref{fig_DiracFloquetBandAnti}(d).

\begin{figure}[tbp]
\scalebox{0.58}{\includegraphics{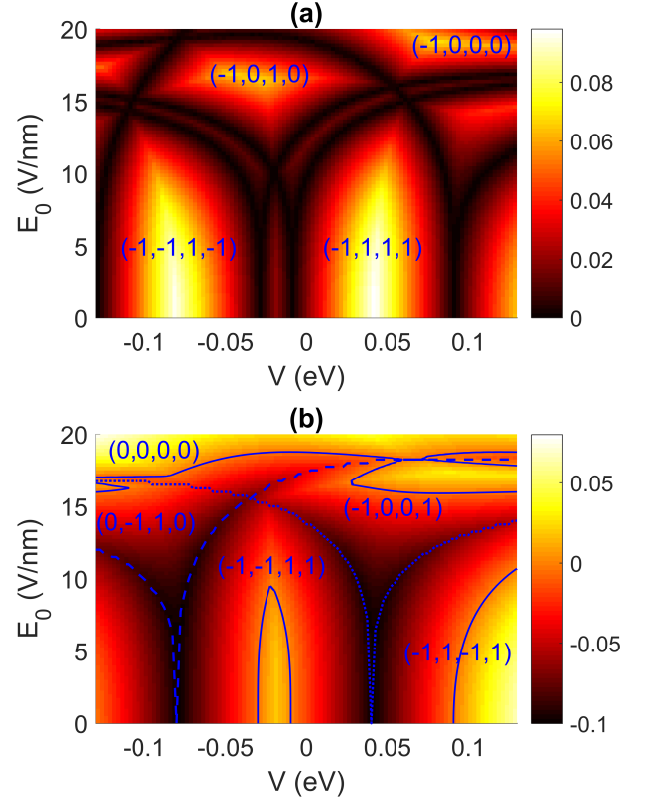}}
\caption{ The band gap of the spin valve in the parameters space of $E_{0}-V$ are plotted as the color scale. The system parameters are $\lambda_{\Delta}^{\pm}=20$ meV and $\lambda_{I}^{\pm}=60$ meV; $\lambda_{AF}^{+}=\lambda_{AF}^{-}=50$ meV in (a), $\lambda_{AF}^{+}=-\lambda_{AF}^{-}=50$ meV in (b). The Chern number of each spin and valley, $(\mathcal{C}_{+}^{+},\mathcal{C}_{+}^{-},\mathcal{C}_{-}^{+},\mathcal{C}_{-}^{-})$ is marked in each phase regime, which is separated by the boundaries where the direct band gap closes. In (b), the boundaries that the band gap equate to zero are marked by solid lines. The boundaries that the direct band gap closes at the band valleys with $s_{z}\tau=1$ or $s_{z}\tau=-1$ are plotted as dashed or dotted lines, respectively. }
\label{fig_gapParallel}
\end{figure}

For the antiferromagnetic spin valve consisting of realistic materials, $\lambda_{\Delta}^{\varsigma}$ and $\lambda_{I}^{\varsigma}$ are nonzero. For example, if the van der Walls spin valve is consisted of BLG being sandwiched between two monolayers of MnPSe$_{3}$, antiferromagnetic exchange field as well as $\lambda_{\Delta}^{\varsigma}$ and $\lambda_{I}^{\varsigma}$ are induced in the graphene layer by the proximity effect. We consider a model with $|\lambda_{AF}^{\varsigma}|=50$ meV, $\lambda_{\Delta}^{\varsigma}=20$ meV and $\lambda_{I}^{\varsigma}=60$ meV to demonstrate the qualitative properties. In this case, the band structure of the two spins and two valleys are different. The band gap of the whole system is given by the overlapping energy range of the band gap of both spins and both valleys. For the spin valve with $\lambda_{AF}^{+}=\pm\lambda_{AF}^{-}$, the band gap in the parameters space of $V-E_{0}$ are plotted in Fig. \ref{fig_gapParallel}(a) and (b), respectively. For the antiparallel exchange field with $\lambda_{AF}^{+}=-\lambda_{AF}^{-}$, the band gap in most part of the phase regime is negative, because the top of valence band is higher than the bottom of the conduction band. Since the direct band gaps of different spin or valley close at different parameters, the combination of the Chern numbers of the two spins and valleys, $(\mathcal{C}_{+}^{+},\mathcal{C}_{+}^{-},\mathcal{C}_{-}^{+},\mathcal{C}_{-}^{-})$, in varying phase regimes are different. The properties of topological phase and edge states in each phase regime are discussed in the next section.

The theoretically predicted Floquet-Bloch band structures in this section can be measured by the vacuum ultraviolet ARPES \cite{YHWang13,Farrell16,YaoWang18,HaifengYang18,BaiqingLv19}, because the optical frequency is assumed to be 7 eV. In our simulation, the maximum amplitude of electric field of the light is 20 $V/nm$, which corresponds to peak power density of $\frac{1}{2}\sqrt{\frac{\varepsilon_{0}}{\mu_{0}}}E_{0}^{2}=0.53$ $W/nm^{2}$. Assuming that a Gaussian beam optical pulse with beam waist being 100 nm and pulse width being 1 ps is applied in experiment, a maximum single pulse energy of $16.7\times10^{-9}$ J is required to scale the phase diagram in our studies.

\section{Tuning of the edge states in zigzag nanoribbon }

In this section, the irradiated BLG zigzag nanoribbons are investigated by applying the tight binding model. The zigzag edges of the nanoribbons are parallel to the $\hat{y}$ axis, so that the zigzag nanoribbon is periodic along the $\hat{y}$ direction, and has $N$ rectangular unit cells along the $\hat{x}$ direction, with eight carbon atoms in each rectangular unit cell. The periodic boundary condition with Bloch phase is applied along the $\hat{y}$ direction. The width of the nanoribbons is $3Na_{C-C}$ with $a_{C-C}=1.42$ ${\AA}$ being the bond length between the carbon atoms.

In principle, the optical field of horizontally incident circular polarized light is described by the vector potential $\mathbf{A}=\hat{x}A_{x}\sin(\mathbf{k}\cdot\mathbf{r}+\Omega t)+\hat{y}A_{y}\sin(\mathbf{k}\cdot\mathbf{r}+\Omega t)+\hat{z}A_{z}\cos(\mathbf{k}\cdot\mathbf{r}+\Omega t)$, where $\mathbf{k}=k_{x}\hat{x}+k_{y}\hat{y}$ with $k_{x}=k\cos(\theta)$, $k_{y}=k\sin(\theta)$, $k=n_{b}\Omega/c$ and $n_{b}$ being the refractive index of the background medium. For the case with $\theta=0$, and then $k_{y}=0$, the optical field is uniform along the $\hat{y}$ direction. Thus, the unit cell is the same as that for non-irradiated BLG zigzag nanoribbon. For the case with $\theta\ne0$, the optical field is nonuniform along the $\hat{y}$ direction, which break the translational symmetry of the nanoribbon. If the wavelength along $\hat{y}$ direction (i.e. $2\pi/k_{y}$) is conformal with the size of the unit cell $\sqrt{3}a_{C-C}$, a supercell is needed to described the system. Since $2\pi/k_{y}$ is much larger than $\sqrt{3}a_{C-C}$, the nonuniform effect could be neglected, so that we apply the approximation that $k_{y}\approx0$. As a result, the optical field is given as $\mathbf{A}=\hat{x}A_{x}\sin(k_{x}x+\Omega t)+\hat{y}A_{y}\sin(k_{x}x+\Omega t)+\hat{z}A_{z}\cos(k_{x}x+\Omega t)$. As a further approximation, $k_{x}$ could be assumed to be zero, which is equivalent to assuming $n_{b}=0$ in the formula. If the background medium has high refractive index, $n_{b}$ is large. Numerical results showed that the band structures of the Floquet states are weakly dependent on $n_{b}$, when $n_{b}$ is smaller than 10, so that only the numerical result with $n_{b}=1$ is shown in this article. In another work, the BLG zigzag nanoribbons are assumed to be suspended in vacuum.

\subsection{Tight Binding Model}

The tight binding model of the BLG is given by the Hamiltonian
\begin{eqnarray}
H=-t\sum_{\langle i,j\rangle}{c_{i}^{\dag}c_{j}}-t_{\perp}\sum_{\langle i,j\rangle_{\perp}}{c_{i}^{\dag}c_{j}}+\sum_{\langle\langle i,j\rangle\rangle}{\lambda_{I}^{\varsigma_{i}}s_{z}\nu_{ij}c_{i}^{\dag}c_{j}}
 \nonumber \\
+\sum_{i}{\lambda_{\Delta}^{\varsigma_{i}}\kappa_{i}c_{i}^{\dag}c_{i}}
+\sum_{i}{\lambda_{AF}^{\varsigma_{i}}s_{z}\kappa_{i}c_{i}^{\dag}c_{i}}
+V\sum_{i}{\varsigma_{i}c_{i}^{\dag}c_{i}}
\end{eqnarray}
, where the first and second summations cover the intra-layer and inter-layer nearest neighbor hopping with strength being $t=2.8$ eV and $t_{\perp}=0.39$ eV, respectively; the third term represents the intrinsic SOC, whose summation cover the next nearest neighbor intra-layer hopping with strength being $\lambda_{I}^{\varsigma_{i}}$; $\varsigma_{i}=\pm1$ represents the top and bottom layer; $\nu_{ij}=\pm1$ for the next nearest neighbor intra-layer hopping with right and left turn. The fourth and fifth terms are the staggered sublattice potential and the antiferromagnetic exchange field, respectively, with $\kappa_{i}=\pm1$ representing A and B sublattices. The sixth term is the energy different between the top and bottom layers due to the gated voltage.

In the presence of the irradiation, the first three terms in the tight binding Hamiltonian is perturbed. The hopping terms include a time-dependent Peierls phase \cite{Peierls33}. For the intra-layer hopping, the Peierls phase is $\gamma(t)=e^{i2\pi\mathbf{A}(t)\cdot\mathbf{r}_{ij}/\Phi_{0}}$, where $\mathbf{r}_{ij}=\mathbf{r}_{i}-\mathbf{r}_{j}$, and $\Phi_{0}=\pi\hbar/e$ is the magnetic flux quantum. For the inter-layer hopping between two sites with the same in-plane coordinations, the Peierls phase is $\gamma_{\perp}(t)=e^{i2\pi A_{z}(t)d_{\perp}(\varsigma_{i}-\varsigma_{j})/(2\Phi_{0})}$, where $d_{\perp}$ is the inter-layer distance. The time dependent factor for the intra-layer and inter-layer hopping terms can be expanded as $e^{iA_{r0}\sin(k_{x}x+\Omega t)}=\sum_{m=-\infty}^{+\infty}{i^{m}J_{m}(A_{r0})e^{-im\pi/2+ik_{x}x+im\Omega t}}$ and $e^{iA_{z0}\cos(k_{x}x+\Omega t)}=\sum_{m=-\infty}^{+\infty}{i^{m}J_{m}(A_{z0})e^{ik_{x}x+im\Omega t}}$, respectively, where $A_{r0}=2\pi(A_{x}\hat{x}+A_{y}\hat{y})\cdot\mathbf{r}_{ij}/\Phi_{0}$ and $A_{z0}=2\pi A_{z}d_{\perp}(\varsigma_{i}-\varsigma_{j})/(2\Phi_{0})$. As a result, the Hamiltonian can be expanded as $H=\sum_{m=-\infty}^{+\infty}{H_{m}e^{im\Omega t}}$. According to the Floquet theorem \cite{Shirley65,Sambe73,Kohler05}, the quantum state can be expressed as $|\Psi_{\alpha}(t)\rangle=e^{i\varepsilon_{\alpha}t/\hbar}\sum_{m=-\infty}^{+\infty}|u_{m}^{\alpha}\rangle e^{im\Omega t}$, with $\varepsilon_{\alpha}$ being the quasi-energy
level of the $\alpha$-th eigenstate and $|u_{m}^{\alpha}\rangle$ the corresponding eigenstate in the m-th Floquet replica. The Floquet states can be obtained by solving the Floquet Hamiltonian, which is defined as $H_{F}=H(t)-i\hbar\frac{\partial}{\partial t}$. The quasi-energy is given by the eigenvalue problem of $H_{F}|\Psi_{\alpha}(t)\rangle=\varepsilon_{\alpha}|\Psi_{\alpha}(t)\rangle$. The Floquet states and the Floquet Hamiltonian can be represented in the Sambe space, which is the direct product space of the Hilbert space and the Fourier space, where the Hilbert space is expanded by the spatial wave function of the eigenstates $\{|u_{m}^{\alpha}\rangle,m\in\mathbb{N}\}$, and the Fourier space is expanded by the series $\{e^{im\Omega t},m\in\mathbb{N}\}$. In this representation, the Floquet state and the Floquet Hamiltonian is time independent. For numerical calculation, the index of the Floquet replica is cutoff at a maximum value as $|m|<m_{max}$. The Floquet Hamiltonian, designated as $\mathcal{H}$, can be represented as a block matrix $\mathcal{H}^{m_{1},m_{2}}$, with the block index $m_{1}$ and $m_{2}$ ranging in $[-m_{max},m_{max}]$. For the diagonal and nondiagonal block, the matrixes are given as $\mathcal{H}^{m_{1},m_{1}}=H_{0}+m\hbar\Omega\mathbf{I}$ and $\mathcal{H}^{m_{1},m_{2}}=H_{m_{1}-m_{2}}$, respectively. Diagonalization of the Floquet Hamiltonian gives the eigenvalue $E_{\alpha}$ and the eigenstate. For the irradiation with high frequency, $m_{max}=2$ is sufficient.

\subsection{BLG without intrinsic SOC}

\begin{figure}[tbp]
\scalebox{0.58}{\includegraphics{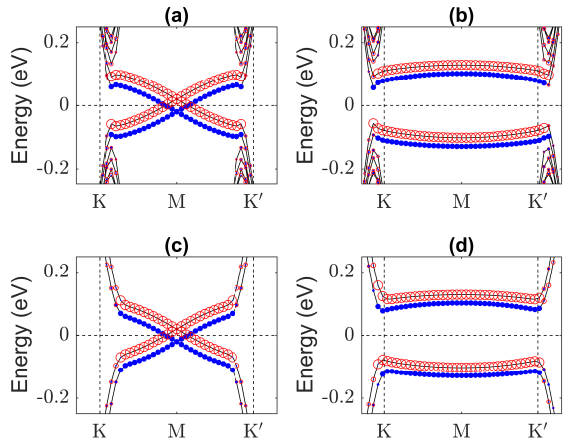}}
\caption{ Band structure of the zigzag nanoribbons with $\Delta_{+}=\Delta_{-}\equiv\Delta_{0}=20$ meV, $\lambda_{\Delta}^{\varsigma}=\lambda_{I}^{\varsigma}=V=0$ and the amplitude of the irradiation being $E_{0}=20$ $V/nm$. Only the band structure of spin up is plotted. The incident angle of the optical field is $\theta=0$ and $\theta=\pi/2$ in (a,c) and (b,d), respectively. The width of the nanoribbon is $N=50$ and $N=10$ in (a,b) and (c,d), respectively. The size of the blue (solid) and red (empty) dots represent the degree of localization at the left and right zigzag terminations, respectively.   }
\label{fig_zigslab0}
\end{figure}

In this subsection, the zigzag nanoribbon with the corresponding bulk state in Fig. \ref{fig_DiracFloquetBand} is studied. Specifically, the band structures of BLG zigzag nanoribbons with $\Delta_{+}=\Delta_{-}\equiv\Delta_{0}=20$ meV, $\lambda_{\Delta}^{\varsigma}=\lambda_{I}^{\varsigma}=V=0$ is plotted in Fig. \ref{fig_zigslab0}. For the corresponding bulk state, the Chern number of each valley is zero, so that no topological edge state is expected. The band structure of the nanoribbon with width $N=50$ and incident angle $\theta=0$ contains two pairs of gapless edge states, as shown in Fig. \ref{fig_zigslab0}(a). At the $M$ point of the Brillouin zone, the pair of edge states localized at the left (right) termination are two-fold degenerated at energy level $\Delta_{0}$ ($-\Delta_{0}$). The degeneration can be broken by turning on the gated voltage $V$, or by changing the incident angle $\theta$. Thus, the edge states are not topological. If the incident angle is turned to $\theta=\pi/2$, the gap of the edge states is maximized, as shown by the band structure in Fig. \ref{fig_zigslab0}(b). As a result, the incident angle can control the band gap of the edge state, which in turn control the conductivity of the nanoribbons. The finite size effect pushes the bulk states into higher energy range, but has small impact on the edge states, as shown in Fig. \ref{fig_zigslab0}(c) and (d), where the width of the nanoribbon is decreased to $N=10$.

\subsection{BLG with intrinsic SOC}

In this subsection, the zigzag nanoribbons with the corresponding bulk state in Fig. \ref{fig_gapParallel}(a) and width $N=50$ are studied.

\begin{figure}[tbp]
\scalebox{0.4}{\includegraphics{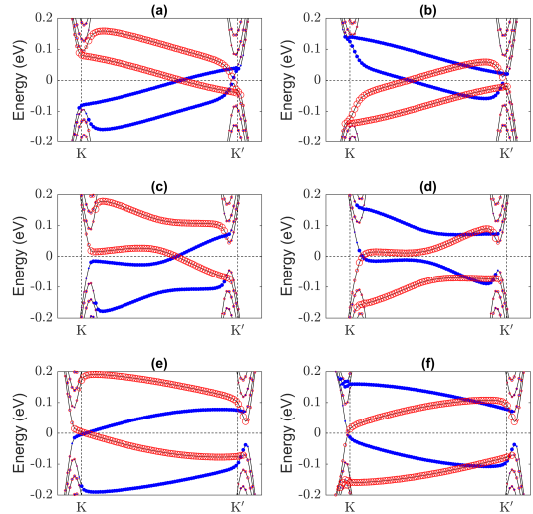}}
\caption{ Band structure of the zigzag nanoribbon with gated voltage being $V=-50$ meV. The band structure of spin up and down electron are plotted in the left and right column, respectively. The amplitude of the irradiation is $E_{0}=0$ in (a,b), and $E_{0}=16.5$ $V/nm$ in (c-f). The incident angle of the optical field is $\theta=0$ and $\theta=\pi/2$ in (c,d) and (e,f), respectively. The other parameters are the same as those in Fig. \ref{fig_gapParallel}(a). The size of the blue (solid) and red (empty) dots represent the degree of localization at the left and right zigzag terminations, respectively.   }
\label{fig_zigslab1}
\end{figure}

For a fixed gated voltage $V=-50$ meV, the bulk band gap and topological phase can be tuned by the amplitude of the irradiation, as shown in Fig. \ref{fig_gapParallel}(a). In the absence of the irradiation, the bulk state is in the phase regime with Chern numbers being $(-1,-1,1,-1)$. This phase is designated as spin-polarized QAH/QVH phase. The band structure of the zigzag nanoribbon in this phase is plotted in Fig. \ref{fig_zigslab1}(a-b). For the spin up electron, the total Chern number is $-2$, so that the bulk state is in the QAH phase. Two pairs of gapless chiral edge states appear in the zigzag nanoribbon, as shown in Fig. \ref{fig_zigslab1}(a). For the spin down electron, the valley Chern number is two, so that the bulk state is in the QVH phase. Two pairs of valley-Hall edge states appears, as shown in Fig. \ref{fig_zigslab1}(b), which is consistence with the absolute value of the Chern number \cite{Thouless82}. The valley-Hall edge states corresponding to K and K$^{\prime}$ valleys are connected with the same valence (conduction) bands, so that the direct band gap is finite \cite{Xintao15}.  As the irradiation with $E_{0}=16.5$ $V/nm$ being turned on, the Chern numbers are switched to be $(-1,0,1,0)$. For each spin, the total Chern number is the same as the valley Chern number. The total Chern numbers of the two spins have opposite sign. One pair of chiral edge states appear in the band structure of the zigzag nanoribbon for each spin, as shown in Fig. \ref{fig_zigslab1}(c-f). The chirality of the chiral edge states of the two spins are opposite to each other.

The band structure of the topological edge states can be tuned by changing the incident angle of the optical field. As $\theta=0$, the bands of the edge state have a segment of flat band near to the Fermi level, as shown in Fig. \ref{fig_zigslab1}(c-d). The band structure has van-Hove singularities, so that the density of state become nearly infinite at the corresponding energy level. As $\theta=\pi/2$, the bands of the edge states become nearly linearly dispersive in a large range of momentum near to the K point, as shown in Fig. \ref{fig_zigslab1}(e-f). In this case, the density of state near to the Fermi level is nearly a constant as of energy. The tuning of the density of state by changing $\theta$ could be measured by the pump-probe setup in experiment \cite{Kitagawa11}. As the frequency of the probe light being resonant to the energy level with van-Hove singularity, the optical absorption could be enhanced \cite{Saroka17,Ivan13}.

\begin{figure}[tbp]
\scalebox{0.4}{\includegraphics{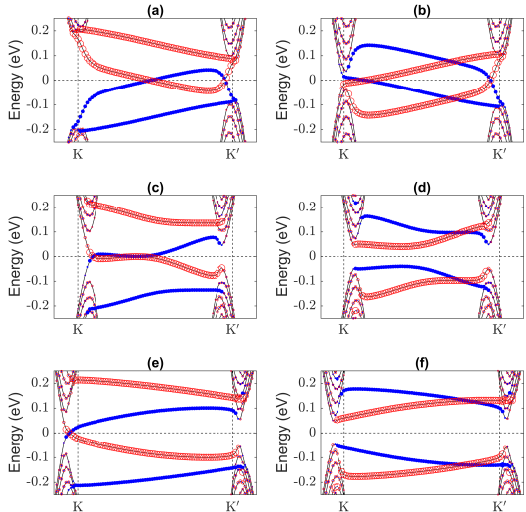}}
\caption{ The same plot as those in Fig. \ref{fig_zigslab1}, except that the gated voltage is $V=77.5$ meV, and the amplitude of the irradiation is $E_{0}=19$ $V/nm$ in (c-f). }
\label{fig_zigslab2}
\end{figure}

If the gated voltage is changed to be $V=77.5$ meV, the corresponding bulk states are in the phase regime with Chern number being $(-1,1,1,1)$ or $(-1,0,0,0)$, as the amplitude of the irradiation equates to $E_{0}=0$ $V/nm$ or $E_{0}=19$ $V/nm$, respectively. Similar to the previous case, in the absence of the irradiation, the bulk state is in the spin-polarized QAH/QVH phase. The corresponding band structure in zigzag nanoribbon are shown in Fig. \ref{fig_zigslab1}(a-b). In the presence of the irradiation, only the K valley of the spin up band have nonzero Chern number. Thus, the corresponding bulk state is in the spin-polarized QAH phase. Only one pair of gapless chiral edge states appear in the band structure of spin up electron in the zigzag nanoribbon, as shown in Fig. \ref{fig_zigslab1}(c-f). As $\theta=0$, the energy level of the van-Hove singularities of the spin up electron is very close to the Fermi level, while that of the spin down electron is near to the edge of the insulating band, as shown in Fig. \ref{fig_zigslab1}(c-d). The van-Hove singularities near to the Fermi level could enhance the spin-polarized carrier concentration, which in turn increase the electrical conductivity. The properties could be applied to improve the performance of opto-spintronic device based on BLG nanoribbon \cite{Chuanxu14}. As $\theta=\pi/2$, the band crossing of the chiral edge states of spin up electron has nearly linear dispersion near to the K point. The tuning of the spin dependent conductivity could be measured in a quantum tunneling photoconductive device \cite{JWMcIver20}.

\section{Conclusion}

In conclusion, the presence of horizontally incident circular polarized optical field could modify the band structure and topological phase of the BLG in antiferromagnetic van der Walls spin valves. The high frequency expansion of the Dirac Fermion model reveals that the irradiation effectively changes the inter-layer hopping, which in turn tunes the physical properties of the Floquet states. The spin-polarized QAH phase with Chern number being one is found, which has one pair of spin-polarized chiral edge states. By changing the incident angle of the optical field, the band gap of the trivial edge states can be tuned; van-Hove singularity could be generated in the dispersion of the chiral edge states.

\begin{acknowledgments}
This project is supported by the National Natural Science Foundation of China (Grant:
11704419).
\end{acknowledgments}

\clearpage

\end{document}